\documentclass[twocolumn,showpacs,pra,groupedaddress,superscriptaddress,nofootinbib,longbibliography]{revtex4-2}
\usepackage{graphicx,amsmath,amsthm,amssymb,amsbsy,subfigure,hyperref,bbm,times,txfonts,float}
\usepackage{subfigure,hyperref,bbm,times}
\usepackage{float}
\usepackage[T1]{fontenc}
\usepackage{braket}
\usepackage{epsfig}
\usepackage{color}
\usepackage{graphicx}% Include figure files
\usepackage{subfigure}
\usepackage{float}
\usepackage{dcolumn}% Align table columns on decimal point
\usepackage{bm}% bold math

\providecommand{\openone}{\leavevmode\hbox{\small1\kern-3.8pt\normalsize1}}

\usepackage{soul}

\hypersetup{
	colorlinks=true,
	linkcolor=blue,
}
\graphicspath{{figure/}}

\begin{document}
	
	%\preprint{APS/123-QED}
	
	\title{Witnessing global memory effects of multiqubit correlated noisy channels by Hilbert-Schmidt speed}% Force line breaks with \\

\author{Kobra Mahdavipour}
\email{kobra.mahdavipour@inrs.ca}
\affiliation{Dipartimento di Ingegneria, Università degli Studi di Palermo, Viale delle Scienze, 90128 Palermo, Italy}
\affiliation{Institut national de la recherche scientifique, Centre Énergie Matériaux et Télécommunications (INRS-EMT), 1650 boulevard Lionel-Boulet, Varennes, Québec J3X 1P7, Canada }
%%%%%%%%%%%%%%%%%%%%%%%%%%%%%%%%%%%%%%%%%%%%%%%%%%%%%%%%%%%%%%%%%%%%%%%%%%%%%%%
\author{Samira Nazifkar}
 \email{Nazifkar@neyshabur.ac.ir}
 \affiliation{Department of Physics, University of Neyshabur, Neyshabur, Iran}
% %%%%%%%%%%%%%%%%%%%%%%%%%%%%%%%%%%%%%%%%%%%%%%%%%%%%%%%%%%%%%%%%%%%%%%%%%%%%%%%%
\author{Hossein Rangani Jahromi}
 \email{h.ranganijahromi@jahromu.ac.ir}
 \affiliation{Physics Department, Faculty of Sciences, Jahrom University, P.B. 74135111, Jahrom, Iran}
%%%%%%%%%%%%%%%%%%%%%%%%%%%%%%%%%%%%%%%%%%%%%%%%%%%%%%%%%%%%%%%%%%%%%%%%%%%%%%%%%%%%%
\author{Rosario Lo Franco}
\email{rosario.lofranco@unipa.it}
\affiliation{Dipartimento di Ingegneria, Università degli Studi di Palermo, Viale delle Scienze, 90128 Palermo, Italy}

	\date{\today}% It is always \today, today,

\begin{abstract}
  In correlated noisy channels, the global memory effects on the dynamics of a quantum system depend on both non-Markovianity of the single noisy channel (intrinsic memory) and classical correlations between multiple uses of the channel itself (correlation-based memory). We show that the Hilbert-Schmidt speed (HSS), a measure of non-Markovianity, serves as a reliable figure of merit for evaluating the role of this correlation-based memory on the global memory effects, for both unital and non-unital channels. The intensity of the correlation-based memory is ruled by a classical correlation strength between consecutive applications of the channel.
  We demonstrate that, for unital noisy channels, increasing the number of qubits of the system significantly weakens the sensitivity of the HSS to this classical correlation strength. Such a pattern indicates that the state evolution of large quantum systems may be less prone to be affected by classical correlations between noisy channels. Moreover, assuming the qubits are affected by independent or classically correlated local non-Markovian unital channels, we observe that, as the number of qubits increases,   the collective behavior of the multiqubit system inhibits the non-Markovian features of the overall system dynamics.
\end{abstract}

\maketitle

\section{Introduction}

The inevitable interaction of quantum systems with their environment creates a noisy channel which induces decoherence, leading to the loss of quantum properties of the system \cite{breuer2002theory,rivas2014quantum}. This phenomenon is well described by the theory of open quantum systems. These system-environment interactions can be either memoryless, known as Markovian regime, or possess memory and involve information backflow, referred to as non-Markovian regime. Memory effects occurring within the non-Markovian regime can mitigate the detrimental effects of environment-induced decoherence. This notion of memory stems from the intrinsic structure of the noisy environment, such as its spectral density, and defines the non-Markovianity of the system dynamics. Non-Markovianity has been identified as a valuable resource for quantum information and communication protocols \cite{PhysRevLett.108.160402, PhysRevA.83.042321, PhysRevLett.109.233601,LoFranco2013Review, laine2014nonlocal,dong2018non,LoFranco2012PRA,Sun2024PRA,Rajabalinia2022PRA,Motavallibashi:21,Xu2013NatComm,Gaikwad2024PRL,Gupta2022PRA,Bylicka2016SciRep,Bylicka2014,jahromi2022remote}. 

On the other hand, in quantum information theory memory effects of quantum channels are identified through multiple uses of the channel on a sequence of quantum systems. 
%In quantum information theory quantum channels with memory are characterised by the existence of correlations between successive applications of the channel on a sequence of quantum systems. 
Channels with memory show classical correlations between multiple uses, while memoryless channels do not exhibit such correlations. This different notion of memory, due to correlations between consecutive applications of a channel, plays a role in protecting and enhancing quantum resources. Specifically, these effects can suppress decoherence \cite{hu2018quantum} and enhance quantum coherence \cite{Hu_2019}, quantum correlations \cite{rameshkumar2023dynamics}, entanglement \cite{PhysRevA.65.050301,PhysRevA.107.022405}, teleportation fidelity \cite{Wang_2020}, quantum Fisher information \cite{hu2020protecting}, and measurement uncertainty \cite{Karpat}.

In correlated non-Markovian channels, the global memory effects encompass  both concepts of memory defined above: (i) the intrinsic memory dictated by the channel's non-Markovianity, and (ii) the memory stemming from classical correlations between consecutive uses of the quantum channel. The study presented in Ref.~\cite{PhysRevA.94.032121} explores this interplay under a dephasing scenario, utilizing two measures of non-Markovianity: trace distance \cite{PhysRevLett.103.210401}, which assesses the backflow of information from the environment to the system, and entanglement-based measures \cite{PhysRevLett.105.050403}. The latter measures reveal that classical correlations between multiple uses of non-Markovian quantum channels amplify the overall dynamical non-Markovianity, while the trace distance measure is not affected by these classical correlations. Importantly, for both measures, correlation effects do not modify the time intervals during which non-Markovianity appears.

Additionally, the impact of channel correlations has been studied using the entanglement-based non-Markovianity indicator and a non-Markovianity witness based on the variation of the volume of accessible states, as detailed in Ref.~\cite{sabale2024facets}. These models, however, are limited to scenarios involving modified Ornstein–Uhlenbeck noise, correlated random telegraph noise \cite{kumar2018non}, and correlated non-Markovian amplitude damping channels \cite{tang2022average}. None of these approaches comprehensively address the critical aspect of classical channel correlation on information backflow, as discussed in Ref.~\cite{PhysRevLett.103.210401}.

The Hilbert-Schmidt speed (HSS) \cite{PhysRevA.102.022221,jahromi2021hilbert,hosseiny2023monitoring,rangani2022searching,hosseiny20technique}, a specialized quantum statistical measure, offers a computationally efficient method to detect non-Markovianity without requiring the diagonalization of the system's density matrix \cite{MahdavipourEntropy}. Recognizing the significance of correlated non-Markovian channels in enhancing the quantum speed limit \cite{PhysRevA.100.052305}, increasing channel capacity, and mitigating noise in quantum error correction \cite{sabale2024facets}, we investigate the global dynamic memory effect in correlated noisy channels. Our focus is on how classical correlation influences non-Markovianity, specifically through the backflow of information, utilizing HSS.

This study generalizes the findings of Ref.~\cite{PhysRevA.94.032121} and examines both unital and non-unital channels with varying noisy spectral densities, including Pauli and depolarizing channels for unital cases and amplitude-damping channels for non-unital cases. Our results indicate that HSS reliably measures these correlations and that classical memory significantly impacts non-Markovianity. Furthermore, we demonstrate that in unital noisy channels with a large number of qubits, the HSS sensitivity to the classical correlation strength between consecutive channel uses is considerably reduced. Initially, we analyze the effect of correlated noisy channel uses on two qubits and subsequently extend our analysis to multiqubit noisy channels, capitalizing on the computational simplicity of HSS.  By examining the behavior of these multi-qubit systems, we illustrate how the collective non-Markovian behavior of the entire system diminishes as the number of qubits, each locally interacting with non-Markovian environments, increases.
\par
The paper is organized as follows: in Sec.~\ref{SecII}, we present the general representation of the quantum channel. The quantifier of the non-Markovian memory effect is discussed in Sec.~\ref{SecIII}. In Sec.~\ref{SecIV}, we examine the two-qubit correlated unital noisy quantum channels under various noisy environments. Further, in Sec.~\ref{SecV}, we investigate the non-unital channel. In Sec.~\ref{VI}, we extend our discussion to multiqubit correlated unital noisy channels. Finally, Sec.~\ref{SecVII} summarizes our results.

\section{Correlated quantum channels}\label{SecII}

Quantum channels are fundamental constructs in quantum information theory, representing the physical processes that convert an initial quantum state into a final one. These channels can be classified into two primary types: memory channels and memoryless channels. Memory channels exhibit correlations between successive uses, while memoryless channels treat each use independently. This memory effect is distinct from non-Markovian memory effects, which arise from temporal correlations in the system's dynamics.

Consider a quantum channel $\Phi$ acting on a system of $n$ qubits, each independently  influenced by a noise described by Kraus operator $K_{i_n}$, and hence given by \cite{PhysRevA.65.050301}
\begin{equation}\label{Eq:General_Case} 
\Phi(\rho)=\sum_{i_1 \cdots i_n}(K_{i_n} \otimes \cdots \otimes K_{i_1}) \rho (K_{i_n}^\dagger \otimes \cdots \otimes K_{i_1}^\dagger),
\end{equation}
A channel is termed unital if it maps the identity operator $\mathbb{I}$ to itself $\Phi(\mathbb{I})=\mathbb{I}$. Non-unital channels, on the other hand, do not preserve the identity operator $\Phi(\mathbb{I}) \neq \mathbb{I}$.

\begin{figure}[!t] 
\centering \includegraphics[width=0.49\textwidth]{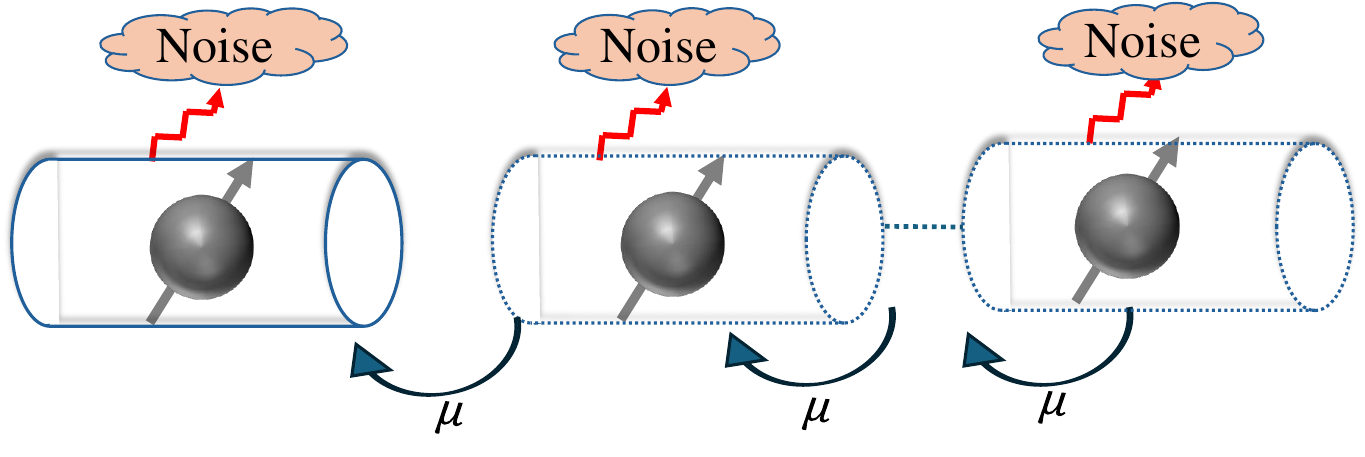} 
\caption{Illustration of the action of correlated noisy channels. The parameter $\mu$ is the classical correlation factor between consecutive uses of a quantum channel on the sequence of quantum systems.} \label{Fig1:correlated_noisy_channel} 
\end{figure}

For unital channels, such as Pauli channels, a generalized framework can be represented by Kraus operators in the following form \cite{PhysRevA.65.050301}
\begin{equation} K_{i_1 \cdots i_n} = \sqrt{p_{i_{1} \cdots i_{n}}} \sigma_{i_1} \cdots \sigma_{i_n}, \end{equation}
where $\sum_{i_1 \cdots i_n} p_{i_1 \cdots i_n} = 1$, $\sigma_0$ denotes the $2 \times 2$ identity matrix, and $\sigma_i$ are the Pauli operators in the $x$, $y$, and $z$ directions. For memoryless channels, the probabilities factorize as $p_{i_1 \cdots i_n} = p_{i_1} p_{i_2} \cdots p_{i_n}$. An interesting extension involves Markov chains, defined as 
\begin{equation} 
p_{i_1 \cdots i_n} = p_{i_1} p_{i_{2}|i_1} \cdots p_{i_{n}|i_{n-1}}, 
\end{equation}
where $p_{i_{n}|i_{n-1}}$ represents the conditional probability that the channel affects the $n$th qubit, given it was applied to the $(n-1)$th qubit. In channels with partial memory, where consecutive uses are correlated, one assumes \cite{PhysRevA.65.050301} 
\begin{equation}
    p_{i_{n}|i_{n-1}} = (1-\mu)p_{i_n} + \mu \delta_{i_{n}, i_{n-1}}.
\end{equation}
Here, $\mu$ is the classical correlation factor (or strength) between consecutive channels, indicating that with probability $\mu$ the same operation is applied to both qubits, while with probability $1-\mu$ the operations are uncorrelated, as depicted in Fig.~\ref{Fig1:correlated_noisy_channel}.

\section{Characterizing Non-Markovianity with Hilbert-Schmidt speed}\label{SecIII}

In this section, we briefly review the definition of the Hilbert-Schmidt speed, previously introduced as a quantifier of the non-Markovian memory effect in open quantum systems.

We begin by introducing the distance measure \cite{gessner2018statistical}
\begin{equation}
	[\text{d}(p,q)]^{2}=\dfrac{1}{2}\sum\limits_{x}^{}|p_{x}-q_{x}|^{2},
\end{equation}
where \( p = \{p_{x}\}_{x} \) and \( q = \{q_{x}\}_{x} \) are the probability distributions. The classical statistical speed can then be defined by quantifying the distance between infinitesimally close distributions taken from a one-parameter family \( ( p_{x} (\phi) ) \) as follows:
\begin{equation}\label{classicalspeed}
	\text{s}\biggl[p(\phi_{0})\biggr]=\dfrac{d}{d\phi}\text{d}\biggl(p(\phi_{0}+\phi),p(\phi_{0})\biggr).
\end{equation}

These classical notions can be extended to the quantum case by considering a pair of quantum states $\rho$ and $\sigma$, and writing $p_{x} = \text{Tr}\{E_{x}\rho\}$ and $ q_{x} = \text{Tr}\{E_{x}\sigma\} $, which represent the measurement probabilities corresponding to the positive-operator-valued measure (POVM) defined by the $ \{E_{x}\geq 0\} $ satisfying $ \sum\limits_{x}^{} E_{x} = \mathbb{I} $.

The corresponding quantum distance, known as the Hilbert-Schmidt distance \cite{OZAWA2000158}, can be obtained by maximizing the classical distance over all possible choices of POVMs \cite{PhysRevA.69.032106}
\begin{equation}\label{qdis}
	\text{D}(\rho,\sigma)\equiv\max_{\{E_{x}\}}\text{d}(p,q)=\sqrt{\frac{1}{2}\text{Tr}{\left[\left(\rho-\sigma\right)^{2}\right]}}.
\end{equation}

Consequently, the Hilbert-Schmidt speed (HSS), the corresponding quantum statistical speed, is introduced as
\begin{equation}\label{quantumspeed}
    \mathrm{HSS}\left(\rho_{\phi}\right)\equiv \mathrm{HSS}{\phi}\equiv\max_{\{E_{x}\}} \text{s}\big[p(\phi)\big]=\sqrt{\frac{1}{2}\text{Tr}\left[\bigg(\dfrac{d\rho_{\phi}}{d\phi}\bigg)^{2}\right]},
\end{equation}
which can be easily computed without the need for diagonalizing the matrix $d\rho_{\phi}/{d\phi}$.

It is now useful to recall the proposed non-Markovianity witness based on the HSS \cite{PhysRevA.102.022221}. We consider an $n$-dimensional quantum system whose initial state is given by
\begin{equation}\label{eq:Eq5}
	|\psi_{0}\rangle=\dfrac{1}{\sqrt{n}}\big(\text{e}^{i\phi}|\psi_{1}\rangle+\ldots+|\psi_{n}\rangle\big),
\end{equation}
where $\phi$ is an unknown phase shift, and $\{|\psi_{1}\rangle,\ldots,|\psi_{n}\rangle\} $ indicates a complete and orthonormal set (basis) for Hilbert space $\mathcal{H}$.
To do so, the HSS-based witness of non-Markovianity is introduced as
\begin{equation}\label{eq:Eq6}
	\chi(t)\equiv \dfrac{d\mathrm{HSS}\big(\rho_{\phi}(t)\big)}{dt} > 0,
\end{equation}
and, consequently, the degree of non-Markovianity is defined as
\begin{equation}
	\mathcal{N}_\textrm{HSS}=\max_{{\varphi,\{|\psi_{1}\rangle,\ldots,|\psi_{n}\rangle\}}} \int\limits_{\chi(t)>0}^{}\chi(t)\text{dt},
\end{equation}
where $\rho_{\phi}(t)$ denotes the evolved state of the system.

\section{Two-qubit Correlated unital noisy quantum channels}\label{SecIV}
In this section, we discuss Pauli and depolarizing channels as examples of unital channels under various noisy environments.

\subsection{Model I: Two-qubit Pauli channels under two different noisy environments}

\subsubsection{Two-qubit Pauli Channels under Colored Pure Dephasing Reservoir}
We investigate the dynamics of two consecutive Pauli channels influenced by colored pure dephasing noise with partial memory on two qubits, forming two-qubit correlated noisy channels. Simplifying Eq.~\eqref{Eq:General_Case}, the two consecutive channel uses with partial memory can be described as follows:
\begin{equation}\label{Eq:Pauli_channel}
	\rho \to \Phi{(\rho)} = \sum_{i,j} p_{i,j} (\sigma_i \otimes \sigma_j) \rho (\sigma_i \otimes \sigma_j),
\end{equation}
where $ p_{i,j} = (1 - \mu) p_i p_j + \mu p_i \delta_{i,j}$.
To analyze the relationship between multiple uses of correlated quantum channels and non-Markovian memory effects, the coefficient $p_{i,j}$ is time-dependent. The Kraus operators governing these dynamics are given by \cite{PhysRevA.70.010304}
\begin{equation}\label{Kraus_Pauli}
	K_i = \sqrt{p_i} \sigma_i, \quad (i=0,1,2,3),
\end{equation}
where $ p_0 = \frac{1}{2}(1 + \eta(\tau)), p_1 = 0, p_2 = 0, p_3 = \frac{1}{2}(1 - \eta(\tau))$, and 
 $\eta(\tau) = e^{-\tau} \left( \frac{\sin(\tau u)}{u} + \cos(\tau u) \right)$,
with $u = \sqrt{16 \nu^2 - 1} $ and $ \tau = \frac{t}{2 \nu}$.
The parameter $\nu$ controls the non-Markovianity degree of the dephasing process that induces dynamical memory effects. Accordingly, the correlated quantum channel in Eq.~\eqref{Eq:Pauli_channel}, describing the dynamical evolution of the open quantum system, is expressed as
\begin{equation}
	\begin{split}
		\Phi{(\rho)} = & p_{0,3} (\sigma_0 \otimes \sigma_3) \rho (\sigma_0 \otimes \sigma_3) + \\
		& p_{3,0} (\sigma_3 \otimes \sigma_0) \rho (\sigma_3 \otimes \sigma_0) + \\
		& p_{0,0} (\sigma_0 \otimes \sigma_0) \rho (\sigma_0 \otimes \sigma_0) + \\
		& p_{3,3} (\sigma_3 \otimes \sigma_3) \rho (\sigma_3 \otimes \sigma_3).
	\end{split}
\end{equation}
The time evolution of the density matrix for two-qubit systems can be easily computed \cite{PhysRevA.94.032121}.
When parametrizing the initial state in the standard computational basis (see Appendix \ref{All Bases}), the Hilbert Schmidt speed (HSS) is given by
 \begin{equation}
 	\begin{split}
 		\mathrm{HSS} = \frac{1}{4} \sqrt{\left((1-\mu) \eta(\tau)^2 + \mu \right)^2 + 2 \eta(\tau)^2}.
 	\end{split}
 \end{equation}
 HSS depends on the effect of classical correlation in multiple uses of the quantum channel, irrespective of the initial basis (refer to Appendix \ref{2Qubit_Pauli_all_basis} for the expressions of the HSS in other relevant bases, whose definition is given in Appendix \ref{All Bases}). 
 
\begin{figure}[!t]
 	\centering
 	\includegraphics[width=0.48\textwidth]{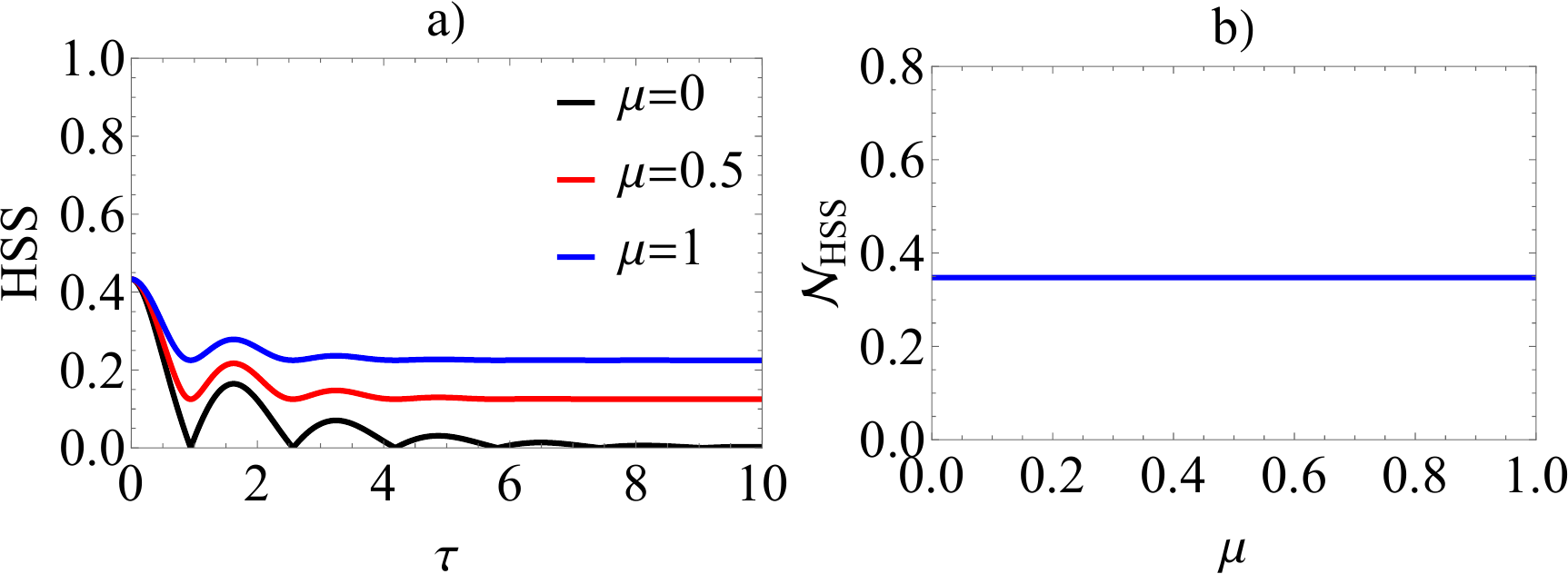}
 	\caption{a) Dynamics of HSS as a function of dimensionless time $\tau=\frac{t}{2\nu}$ for two-qubit Pauli channels with memory under the colored pure dephasing evolution in the standard basis with $\nu=1$. Three different values of the correlation factor $\mu$ are considered. b) HSS-based measure $\mathcal{N}_\mathrm{HSS}$ as a function of the correlation factor $\mu$.}
 	\label{2Pauli-colored}
 \end{figure}

The dynamics of HSS as a function of the correlation coefficient $\mu$ is shown in Fig.~\ref{2Pauli-colored}{a}. The curves demonstrate that HSS, as a bona-fide witness, can evaluate the global memory effect of correlated non-Markovian channels. It serves as both a witness of non-Markovianity and an indicator for monitoring classical memory. An increase in classical correlation between multiple uses of the channel (thus increasing the memory) corresponds to an amplification of the HSS.
 
Notably, increasing the memory of the quantum channel $\mu$ is also associated with an attenuation of the time fluctuations of the HSS. To examine the effect of classical correlation between multiple uses of the channel on the backflow of information, we consider two factors: the time intervals of non-Markovianity and the degree of non-Markovianity. 
 
From the dynamics of HSS, it is apparent that the time intervals for the appearance of non-Markovianity do not change with increasing classical correlation between sequences of noisy Pauli channels. Additionally, we compute the HSS-based measure of non-Markovianity $\mathcal{N}_\textrm{HSS}$ through numerical optimization over numerous initial states. As displayed in Fig.~\ref{2Pauli-colored}{b}, this measure, which quantifies the strength of the backflow of information from the environment to the system, does not increase with the degree of classical correlation between consecutive uses of the channel. 
This finding aligns with the results presented in Ref.~\cite{PhysRevA.94.032121}, which investigates the same scenario using a trace-distance based (BLP) measure of non-Markovianity. This consistency corroborates the complete similarity between the HSS-based and BLP measures of non-Markovianity, even in correlated channels with memory.

\subsubsection{Two-Qubit Pauli Channels under Squeezed Vacuum Reservoir}

\begin{figure}[!t]
	\centering
	\includegraphics[width=0.48\textwidth]{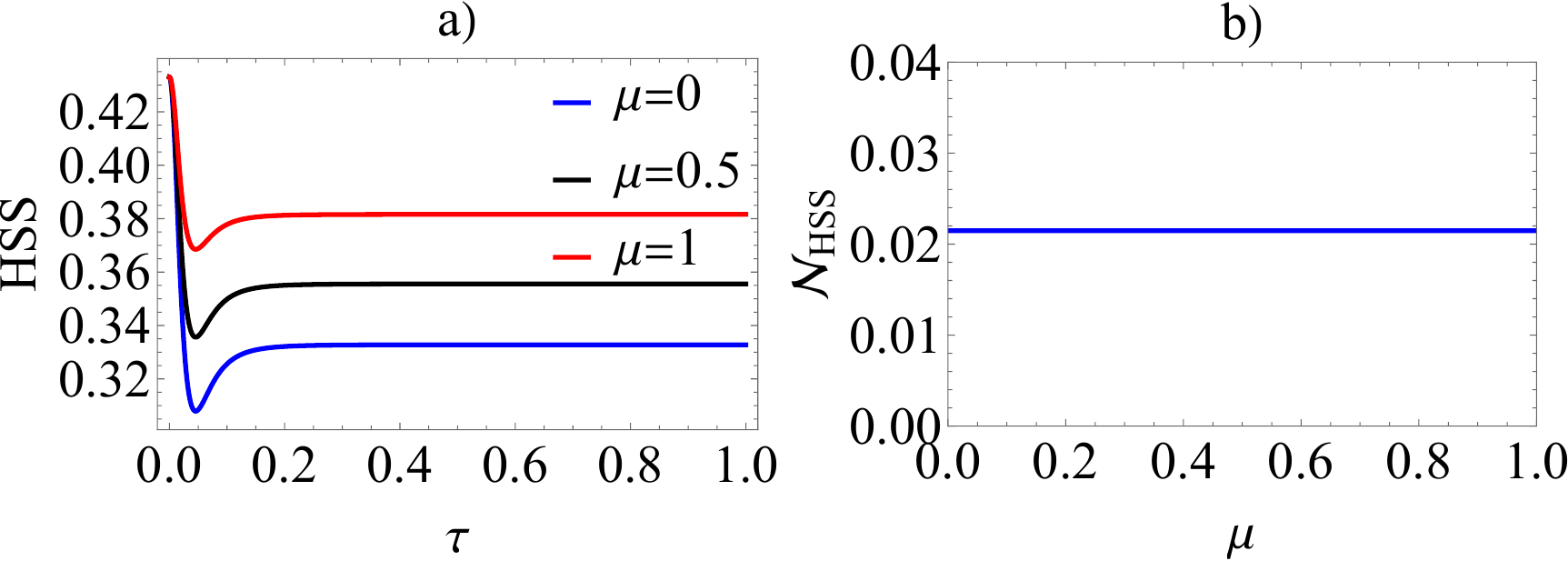}
	\caption{a) The dynamic of HSS as a function of dimensionless time $\tau=\omega t$ for a two-qubit Pauli channel with memory under squeezed vacuum reservoir with $\alpha=0.5$, $s=4$, $r=0.5$ and $\theta=\frac{3\pi}{2}$. b) HSS-based measure $\mathcal{N}_\mathrm{HSS}$ as a function of the correlation factor $\mu$.}
	\label{2Pauli-Squeezed}
\end{figure}

We examine a scenario where two spin-$s$ systems, specifically two-qubit Pauli channels $(s=\frac{1}{2})$, are subjected to a squeezed vacuum reservoir. These two-qubit Pauli channels undergo dephasing and experience decoherence without dissipation. The Kraus operators for such system are given by Eq.~\eqref{Kraus_Pauli}, with
\begin{equation}
    p_0 = \frac{1}{2} \left( e^{-\gamma (t)} + 1 \right), p_1 =p_2 = 0,  p_3 = \frac{1}{2} \left(1 - e^{-\gamma (t)} \right),
\end{equation}
where $\gamma(t)$ denotes the dephasing function. For a squeezed vacuum reservoir characterized by the squeezing parameter \( r \) and the squeezing angle \( \theta \), the dephasing function is defined as follows \cite{JI2020164088}:
\begin{equation}\label{Squeezdec}
	\begin{split}
		\gamma(t) &= \int_0^{\infty} J(\omega) \frac{1 - \cos{\left(\omega t\right)}}{\omega^2} \\
		& \quad \times \left[ \cosh(2r) - \sinh(2r) \cos{(\omega t - \theta)} \right] d\omega,
	\end{split}
\end{equation}
where $J(\omega)$  is the spectral density of the reservoir. For Ohmic-like reservoirs, the spectral density is given by
\begin{equation}
	J(\omega) = \alpha \frac{\omega^s}{\omega_c^{s-1}} \exp\left(-\frac{\omega}{\omega_c}\right),
\end{equation}
in which $\alpha$ represents a dimensionless coupling strength, $\omega_c$ denotes the cutoff frequency of the bath, and the parameter $s$ is positive and characterizes the environmental properties. By varying the Ohmic parameter $s$, one can distinguish between sub-Ohmic ($0 < s < 1$), Ohmic ($s = 1$), and super-Ohmic ($s > 1$) reservoirs.
Using algebraic manipulations and assuming $\omega_c = 20 \omega$, we obtain the dephasing factor $\gamma(\tau)$ in terms of the dimensionless time $\tau = \omega t$ as follows: \cite{Wu2017}
\begin{equation}
	\begin{split}
		\gamma (\tau) =& \frac{1}{2} \alpha \bigg( \cosh(2r) \left[ -(1-20 i\tau )^{1-s} - (1+20i\tau )^{1-s} + 2 \right]+\\
		& \sinh(2r) \cos(\theta) \left[ -2(1-20i\tau )^{1-s} + (1-40i \tau)^{1-s} + 1 \right] \bigg) \\
		&\times \Gamma(s-1),
	\end{split}
\end{equation}
where $\Gamma(.)$ is the Euler Gamma function.

As a consequence, for the two-qubit Pauli channel with classical correlation under a squeezed vacuum reservoir, the amount of Hilbert-Schmidt speed on the standard basis is calculated as
\begin{equation}
	\begin{split}
		\mathrm{HSS} = \frac{1}{4} \sqrt{(\mu -1)^2 e^{-4 \gamma (\tau)} + 2(1 - \mu^2 + \mu) e^{-2 \gamma (\tau)} + \mu ^2},
	\end{split}
\end{equation}

We find that the HSS is sensitive to the presence of classical correlation, irrespective of the basis (see Appendix\ref{2Qubit_Pauli_all_basis} for other bases), and hence it is capable of detecting global memory effects. The dynamics of the HSS are represented in Fig.~\ref{2Pauli-Squeezed}{a)}. Similar to the previous example, as the memory coefficient $\mu$ increases, the HSS enhances. The presence of classical correlation does not alter the time intervals during which temporary backflow of information occurs.

Additionally, we display the degree of non-Markovianity, $\mathcal{N}_{\text{HSS}}$, in Fig.~\ref{2Pauli-Squeezed}{b)}. The degree of non-Markovianity remains constant and is not affected by the correlation factor of Pauli channels under a squeezed vacuum reservoir. This indicates that while classical correlation enhances the HSS, it does not impact the overall structure of non-Markovianity in the system.

 \subsection{Model II: Two-Qubit Depolarizing Channels under Colored Noise Reservoir}
 
 Here, we study an example of a completely positive, trace-preserving map resulting from a system-environment coupling that is fundamentally non-Markovian and defines a depolarizing channel with colored noise \cite{nielsen2010quantum}. The depolarizing channel, an important model for describing noise in quantum systems, can be represented by the Kraus operators given by $ K_i = \sqrt{p_i} \sigma_i$, where $p_i$ are the non-negative linear combinations defined as \cite{PhysRevA.70.010304}
  \begin{equation} \label{pi ge}
 \begin{split}
        p_0 =& \frac{1}{4} (\Lambda_1 + \Lambda_2 + \Lambda_3 + 1), \nonumber \\
 	p_1 =& \frac{1}{4} (\Lambda_1 - \Lambda_2 - \Lambda_3 + 1), \nonumber \\
 	p_2 =& \frac{1}{4} (-\Lambda_1 + \Lambda_2 - \Lambda_3 + 1), \nonumber \\
 	p_3 =& \frac{1}{4} (-\Lambda_1 - \Lambda_2 + \Lambda_3 + 1).
 \end{split}
 \end{equation}
In terms of the dimensionless time $\tau = \frac{t}{2 \nu}$, the functions $\Lambda_i = e^{-\tau} \left( \frac{\sin (\tau \Omega_i)}{\Omega_i} + \cos (\tau \Omega_i) \right)$ are damped harmonic oscillators with frequencies $\Omega_i = \sqrt{(4 \theta_i)^2 - 1}$. We choose $\theta_1 = \theta_2 = \theta_3 = \theta$, under the condition that $\theta$ is in the range $[0, \frac{1}{4}\sqrt{1 + (\frac{\pi}{\log{3}})^2}] $, which results in $\Lambda_1 = \Lambda_2 = \Lambda_3 = \Lambda$.
 
The corresponding Kraus operators are given by $ K_i = \sqrt{\frac{1 - \Lambda}{2}} \sigma_i $ for $ i = 1, 2, 3 $, and $K_0 = \sqrt{\frac{1 + 3\Lambda}{4}} \sigma_0 $.  Thus, the amount of HSS for a two-qubit depolarizing channel with classical correlation under the evolution of colored noise in the standard basis for $\phi = \pi$ is obtained as 
 \begin{equation}
 	\mathrm{HSS} = \frac{1}{4} \sqrt{\Lambda^2 \left(2 (\Lambda (-\mu) + \Lambda + \mu)^2 + 1 \right)}.
 \end{equation}
 
\begin{figure}[!t]
    \centering
 	\includegraphics[width=0.48\textwidth]{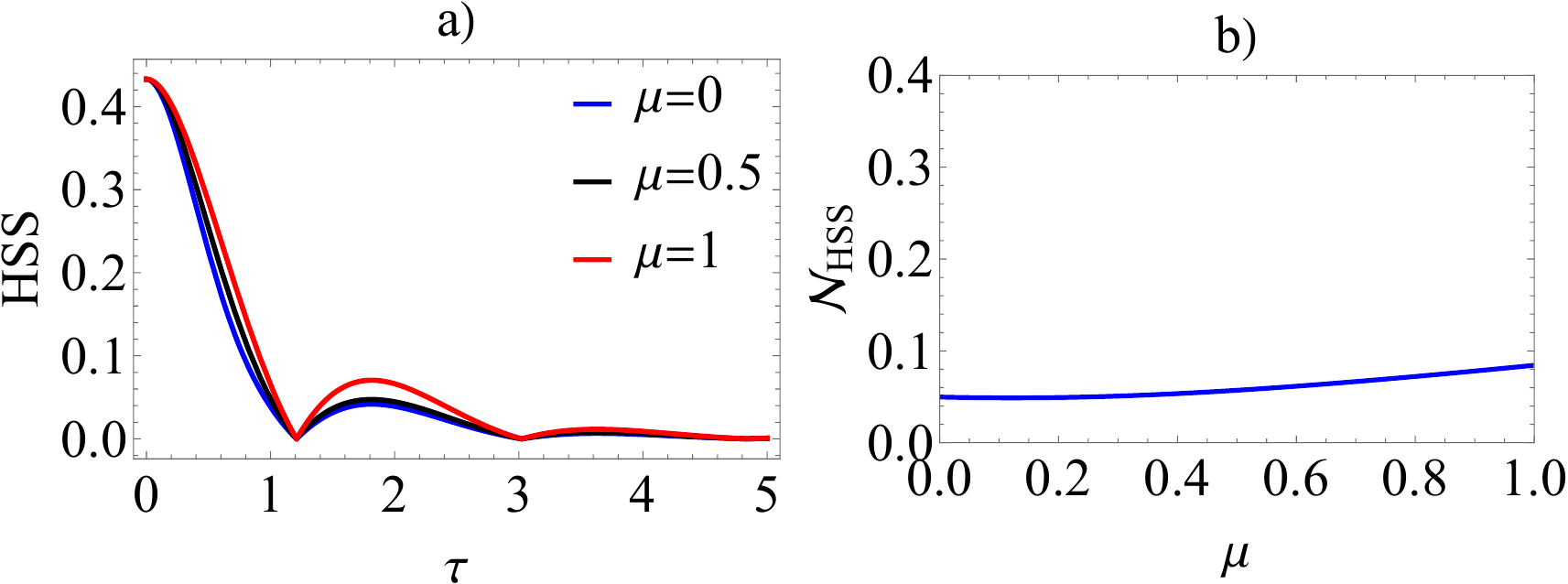}
 	\caption{a) Dynamics of HSS  as a function of dimensionless time $\tau=\frac{t}{2\nu}$ for a two-qubit depolarizing channel with memory under the evolution of colored noise, for $\theta = 0.5$ and $\phi = \pi$. b) HSS-based measure $\mathcal{N}_\mathrm{HSS}$ as a function of the correlation factor $\mu$.}
 	\label{2Depolorizing_channel}
\end{figure}

It is worth mentioning that, similar to previous examples, HSS serves as a faithful witness of the global memory effect in this context. We demonstrate the dynamics of the HSS in Fig.~\ref{2Depolorizing_channel}{a)}, exhibiting behavior analogous to the previous example. The impact of classical correlation is evident in the amplification of HSS. Importantly, it does not affect the number of oscillations or the time intervals in which the non-Markovian signature appears.

Furthermore, we calculate the HSS-based measure of non-Markovianity through numerical optimization over various initial states and their parameterizations, as shown in Fig.~\ref{2Depolorizing_channel}{b)}. The degree of non-Markovianity is observed to be partially enhanced by increasing the memory coefficient between successive uses of quantum channels. This enhancement originates from the increased time fluctuations of the HSS with a rise in the classical correlation between multiple channel uses.

The findings underscore the robustness of HSS in detecting temporal backflow of information and global memory effects in quantum systems. By providing a reliable measure of non-Markovianity, HSS offers valuable insights into the dynamics of quantum information processes under the influence of colored noise and classical correlations.

\section{Two-Qubit Correlated Non-Unital Noisy Quantum Channel}
\label{SecV}

%\textit{Two-qubit amplitude damping channel under Lorentzian reservoir}

In this section, we examine the amplitude damping channel as an example of a non-unital channel. 

The amplitude damping channel describes the dynamics of a qubit interacting with a dissipative environment, associated with spontaneous emission. The time-dependent Kraus operators for the single-qubit amplitude damping channel under a Lorentzian reservoir are given by \cite{Bylicka2014}
\begin{equation}
	K_0 = \left( \begin{array}{cc}
		1 & 0 \\
		0 & G(\tau) \\
	\end{array} \right) , \quad
	K_1 = \left( \begin{array}{cc}
		0 & \sqrt{1 - G(\tau)^2} \\
		0 & 0 \\
	\end{array} \right),
\end{equation}
where $G(\tau)^2 $ represents the decay of the excited population and is given by
\begin{equation} \label{G}
	G(\tau) = e^{-\tau /2} \left( \frac{\sinh \left( \sqrt{\frac{1}{4} - \frac{a}{2}} \tau \right)}{\sqrt{1 - 2a}} + \cosh \left( \sqrt{\frac{1}{4} - \frac{a}{2}} \tau \right) \right).
\end{equation}
Here, $a = \frac{\gamma_0}{\lambda}$ and $ \tau = \lambda t$, with $\gamma_0$ and $\lambda$ describing the coupling strength and the spectral width, respectively. In the weak coupling regime $( \gamma_0 < \frac{\lambda}{2} )$, $|G(\tau)|$ decreases monotonically. In the strong coupling regime $(\gamma_0 > \frac{\lambda}{2}, a > \frac{1}{2} )$, $|G(\tau)|$ oscillates, showing nonmonotonic behavior.

\begin{figure}[t!] 
 \centering \includegraphics[width=0.48\textwidth]{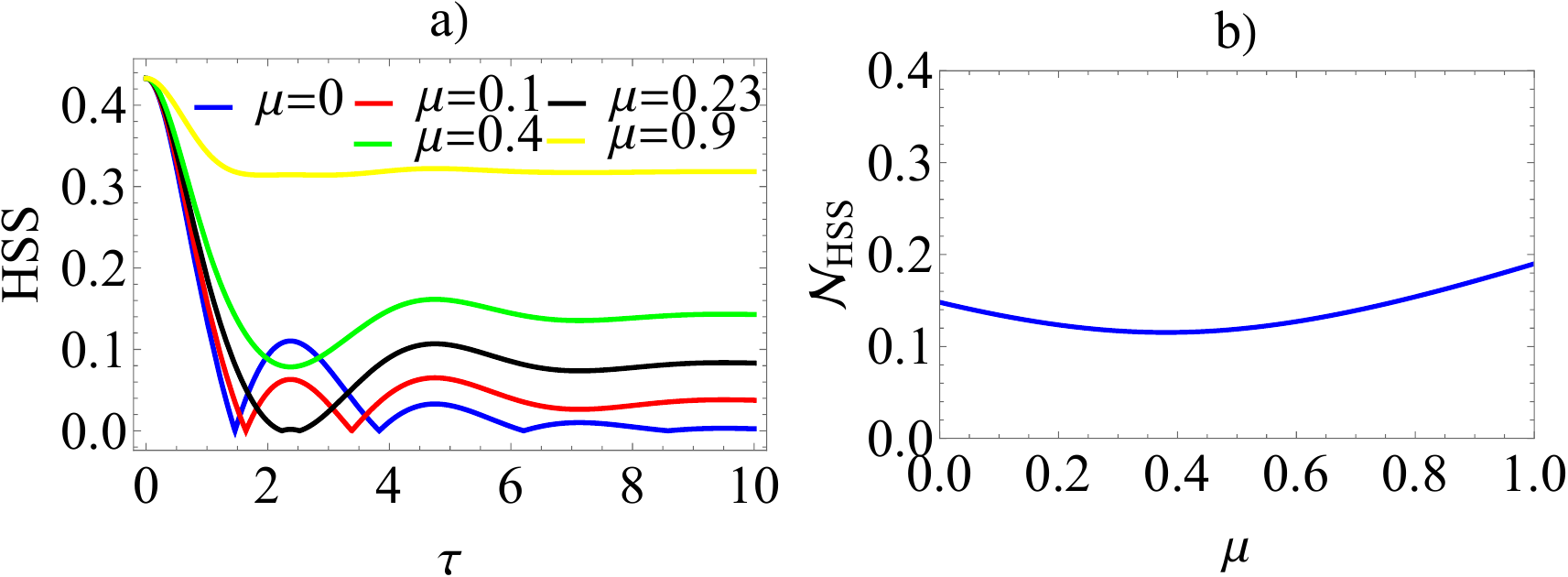} \caption{a) Dynamics of HSS as a function of dimensionless time $\tau=\lambda t$ for two-qubit amplitude damping channels with memory in the standard basis under the evolution of a Lorentzian reservoir for $a=4$ and $\phi=\pi$. b) HSS-based measure $\mathcal{N}_\mathrm{HSS}$ as a function of the correlation factor $\mu$.} \label{2Amplitude damping channels} 
 \end{figure}

The evolution of two sequences of amplitude-damping channels with classical memory is expressed as \cite{PhysRevA.67.064301}
\begin{equation}
	\rho \to \Phi(\rho) = (1 - \mu) \sum_{i,j = 0}^{1} K_{i,j} \rho K_{i,j}^{\dagger} + \mu \sum_{l = 0}^{1} F_{l} \rho F_{l}^{\dagger},
\end{equation}
in which $K_{i,j} = K_i \otimes K_j$, and the Kraus operators for the correlated part, obtained by solving the correlated Lindblad equation, are given by \cite{Huang2017}
\begin{equation}
	F_0 = \left( \begin{array}{cccc}
		1 & 0 & 0 & 0 \\
		0 & 1 & 0 & 0 \\
		0 & 0 & 1 & 0 \\
		0 & 0 & 0 & G(\tau) \\
	\end{array} \right),
	F_1 = \left( \begin{array}{cccc}
		0 & 0 & 0 & \sqrt{1 - G^2(\tau)} \\
		0 & 0 & 0 & 0 \\
		0 & 0 & 0 & 0 \\
		0 & 0 & 0 & 0 \\
	\end{array} \right).
\end{equation}
For the two-qubit amplitude damping channel with classical correlation under a Lorentzian reservoir, the Hilbert-Schmidt speed (HSS) on the standard basis is obtained as
\begin{equation}
	\mathrm{HSS} = \frac{1}{4} \sqrt{\left( G(\tau)^2 + 2 \right) \left( G(\tau) + \mu - \mu G(\tau) \right)^2}.
\end{equation}
Similar to the behavior observed in non-dissipative (unital) channels, the HSS is shown to be an effective tool for detecting global memory effects even in dissipative (non-unital) channels. As illustrated in Fig~.\ref{2Amplitude damping channels}{a)}, the dynamics of HSS for a correlated amplitude damping channel, a typical non-unital channel, exhibit distinct characteristics when compared to unital channels, particularly as the memory parameter of the channel increases. Consequently, the time interval during which non-Markovian effects re-emerge shifts, highlighting a dependency on the memory parameter. In more detail, when starting from a special initial state, the HSS fluctuations over time are notably weakend as classical correlations increase. To further investigate the impact of these classical correlations on the strength of non-Markovianity, we computed the HSS-based measure of non-Markovianity through optimization over initial bases. The results, presented in Fig.~\ref{2Amplitude damping channels}{b)}, demonstrate that as the correlation factor increases, the strength of non-Markovianity initially decreases, followed by an increase.

Accordingly, the HSS provides a nuanced quantification of quantum state dynamics, revealing intricate details about the temporal evolution of quantum systems. In non-unital channels, where dissipative effects play a crucial role, HSS acts as a precise indicator of memory retention and system coherence. This capability is essential in distinguishing between different types of quantum noise and their impacts on system dynamics.

The weakened HSS fluctuations with increasing classical correlations, starting from a specific initial state, imply that these correlations play a stabilizing role, effectively shielding the system from environmental disturbances. This stabilization can be attributed to the decoherence mitigation provided by stronger correlations, which enhance the resilience of the quantum state against external noise.

Furthermore, the non-monotonic relationship between the correlation factor and the strength of non-Markovianity, as derived from the HSS-based measure, suggests a complex interplay between memory effects and system-environment interactions. Initially, the increasing classical correlations reduce the extent of non-Markovian behavior by enhancing the system’s coherence. However, as the correlation factor continues to rise, the system starts exhibiting increased memory retention, leading to a resurgence of non-Markovian dynamics. This resurgence implies that beyond a certain threshold, classical correlations not only preserve the quantum state for a specific evolution but also enhance its ability to retain information about its past interactions.

\section{Multiqubit correlated unital noisy quantum channel}\label{VI}

In this study, we extend our analysis beyond the classical correlations observed in consecutive applications of noisy channels on two-qubit systems. We explore scenarios where noisy channels operate multiple times on sequences of multiqubit systems, with a particular focus on unital channels, such as multiqubit correlated Pauli and depolarizing noisy channels. To investigate the memory effects of noisy channels on high-qubit systems, we calculate the HSS for systems with up to eight qubits, utilizing the unital noisy channel on a standard basis. Given the complexity of the analytical expressions for the HSS in high-dimensional systems, our analysis prioritizes the interpretation of the figures rather than presenting the explicit formulas.

\begin{figure}[!t] 
\centering \includegraphics[width=0.48\textwidth]{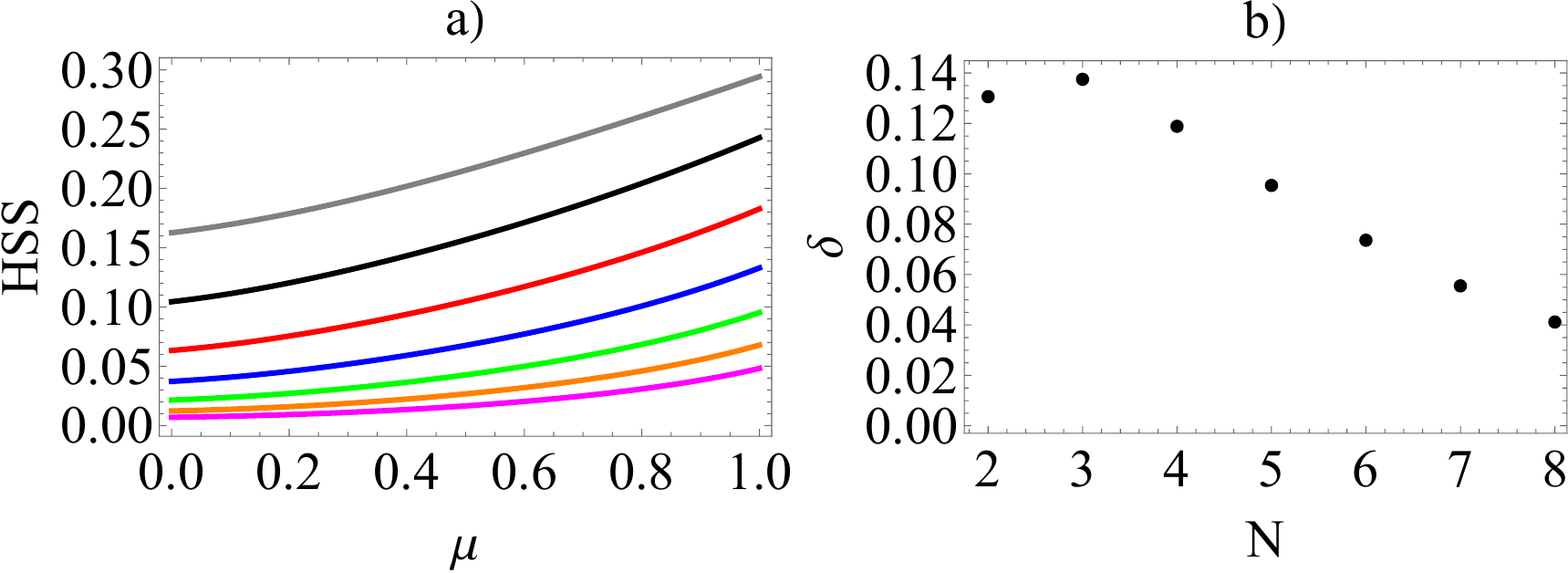} \caption{a) Dynamics of HSS for multiqubit correlated Pauli channel under colored pure dephasing reservoirs with constant reservoir parameters $\nu=1$ at dimensionless time $\tau=1.62$. The solid gray, black, red, blue, green, orange, and magenta lines correspond to $N$ = 2, 3, 4, 6, 7, and 8 qubits. b) Range of variation $\delta$ versus the number of qubits.} \label{Multiqubit-Pauli-colored} 
\end{figure}

\begin{figure}[!h] 
\centering \includegraphics[width=0.48\textwidth]{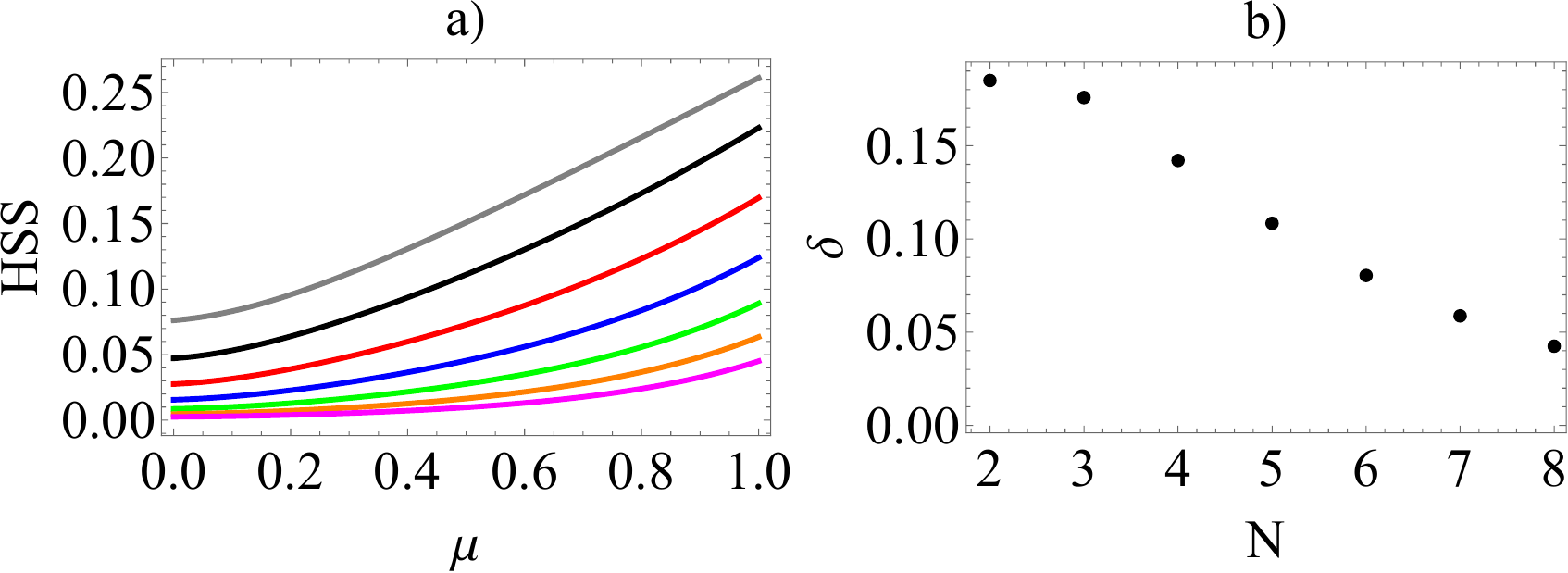} \caption{a) Dynamics of HSS for multiqubit correlated Pauli channel under squeezed vacuum reservoirs: $\alpha=0.5$, $s=4$, $r=0.5$, and $\theta=\frac{3\pi}{2}$ at dimensionless time $\tau=0.2$. The solid gray, black, red, blue, green, orange, and magenta lines correspond to $N$ = 2, 3, 4, 6, 7, and 8 qubits. b) Range of variation $\delta$ versus the number of qubits.} \label{Multiqubit-Pauli-Squeezed} 
\end{figure}

\begin{figure}[!t] 
\centering \includegraphics[width=0.48\textwidth]{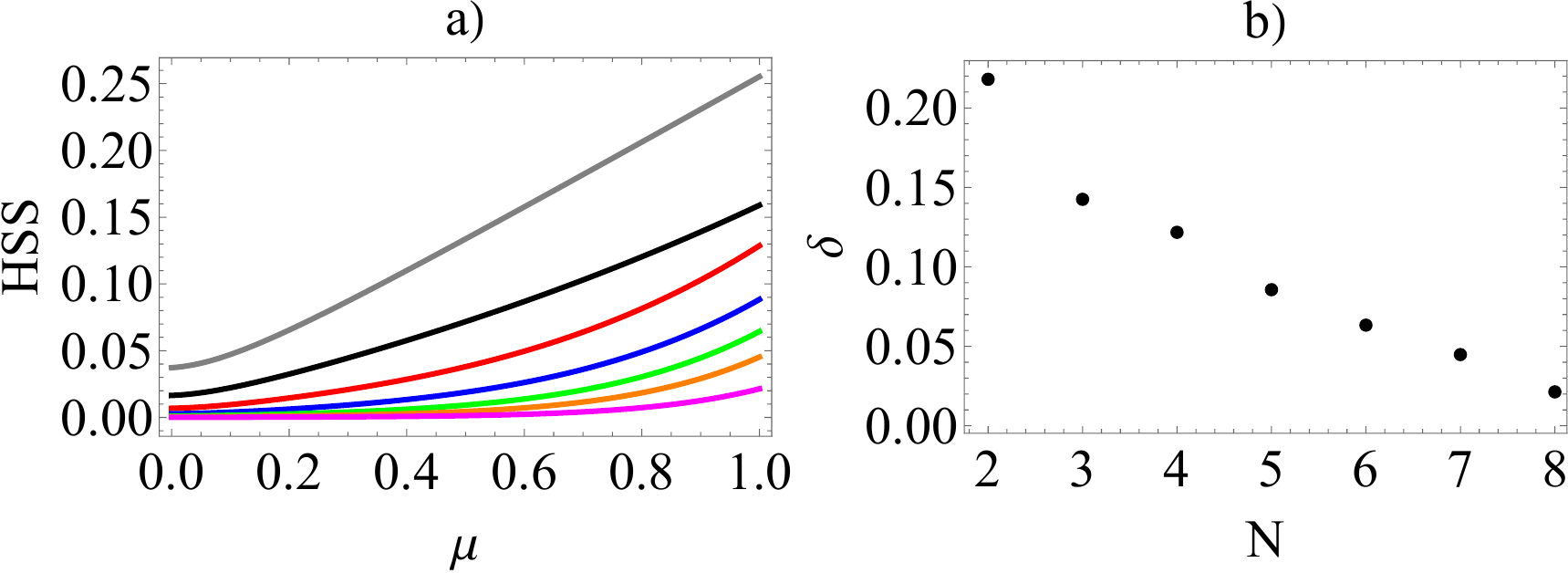} 
\caption{a) Dynamics of HSS for multiqubit correlated depolarizing channel under colored noise reservoirs: $\eta=0.5$, $\Phi=\frac{\pi}{2}$ at dimensionless time $\tau=1.6$. The solid gray, black, red, blue, green, orange, and magenta lines correspond to $N$ = 2, 3, 4, 6, 7, and 8 qubits. b) Range of variation $\delta$ versus the number of qubits.} \label{Multiqubit-depolarizing} 
\end{figure}

The HSS dynamics under constant reservoir parameters at given times are illustrated for the Pauli channel under a colored pure dephasing reservoir, the Pauli channel under a squeezed vacuum reservoir, and the depolarizing channel under a colored noise reservoir, as shown in Fig.~\ref{Multiqubit-Pauli-colored}{a)}, Fig.~\ref{Multiqubit-Pauli-Squeezed}{a)}, and Fig.~\ref{Multiqubit-depolarizing}{a)}, respectively.

\begin{figure}[t]
	\centering
	\includegraphics[width=0.48\textwidth]{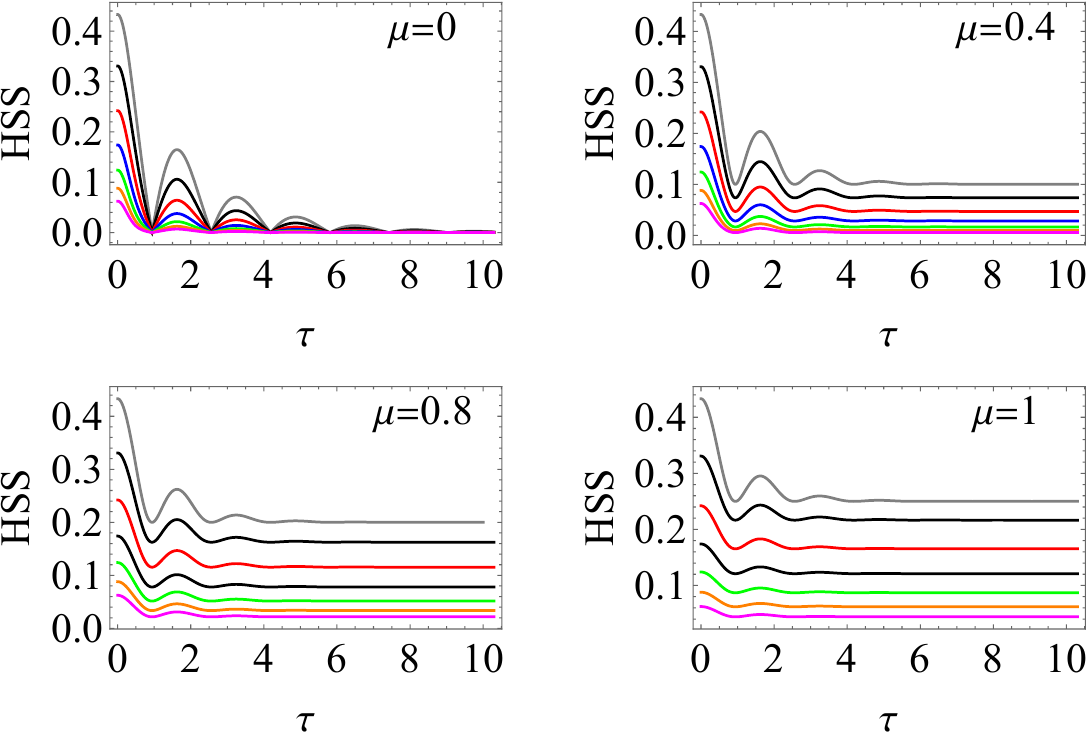}
	\caption{Dynamics of HSS as a function of dimensionless time $\tau=\frac{t}{2\nu}$ for correlated multi-qubit Pauli channels under colored pure dephasing evolution in the standard basis with $\nu=1$. The solid gray, black, red, blue, green, orange, and magenta lines correspond to $N = 2, 3, 4, 6, 7$, and $8$ qubits.}
	\label{HSS_multiqubit_Pauli_pure_dephasing}
\end{figure}

\begin{figure}[t]
	\centering
	\includegraphics[width=0.48\textwidth]{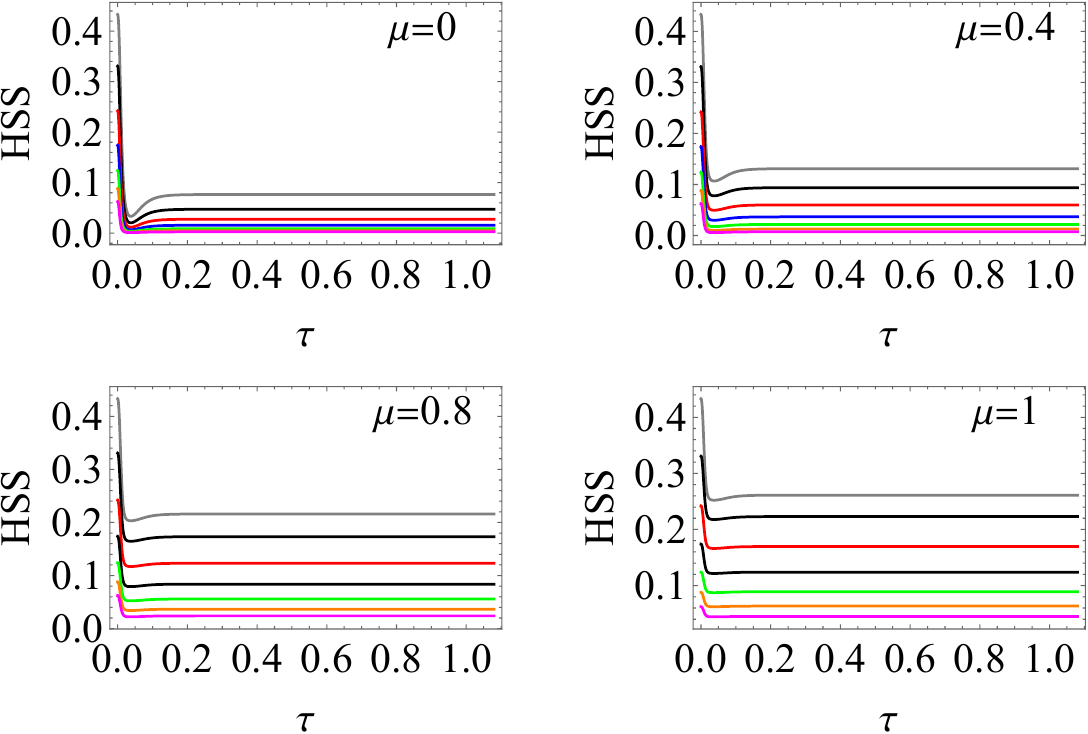}
	\caption{Dynamics of HSS as a function of dimensionless time $\tau=\omega t$ for correlated multi-qubit Pauli channels under squeezed vacuum evolution in the standard basis with $\alpha=0.5$, $s=4$, $r=0.5$, and $\theta=\frac{3\pi}{2}$. The solid gray, black, red, blue, green, orange, and magenta lines correspond to $N = 2, 3, 4, 6, 7$, and $8$ qubits.}
	\label{HSS_multiqubit_Pauli_squuezed}
\end{figure}

\begin{figure}[t]
	\centering
	\includegraphics[width=0.48\textwidth]{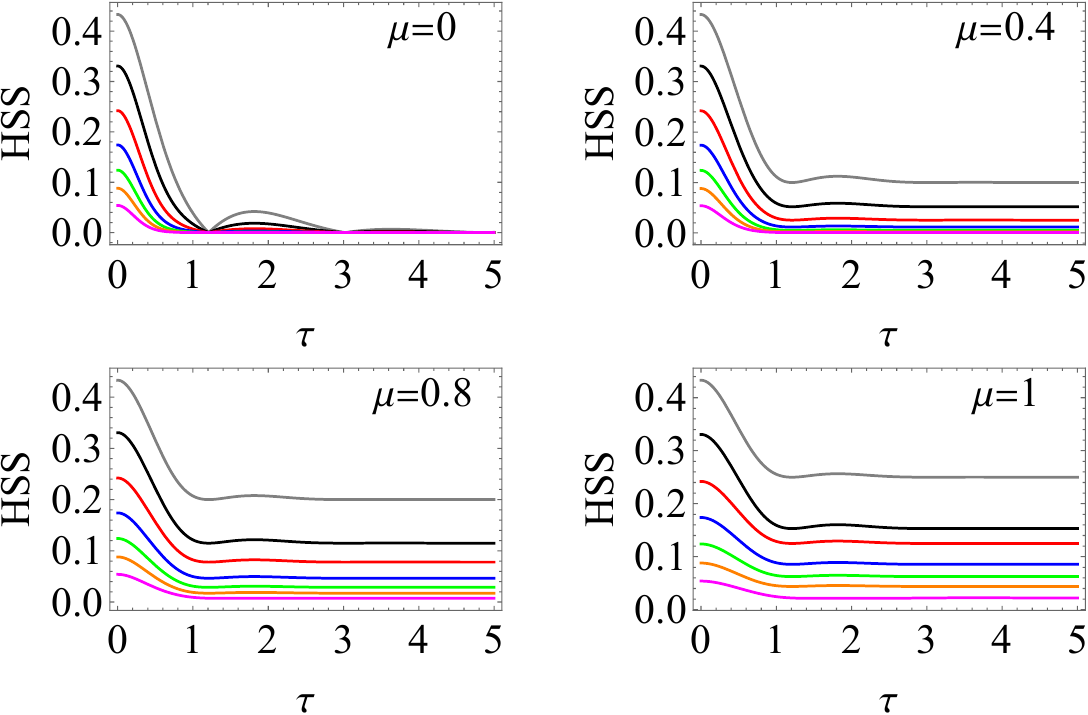}
	\caption{Dynamics of HSS as a function of dimensionless time $\tau=\frac{t}{2\nu}$ for correlated multi-qubit depolarizing channels under colored noise evolution in the standard basis with $\theta=0.5$ and $\phi=\frac{\pi}{2}$. The solid gray, black, red, blue, green, orange, and magenta lines correspond to $N = 2, 3, 4, 6, 7$, and $8$ qubits.}
	\label{HSS_multiqubit_depolarizing}
\end{figure}

Across all scenarios, regardless of the number of qubits, increasing the classical correlation typically amplifies the dynamics of the HSS. However, as demonstrated, this trend does not hold for a high number of qubits. In these instances, increasing the classical correlation does not significantly amplify or enhance the dynamics of the HSS. In other words, the effect of the presence of classical correlation $\mu$ between multiple uses of quantum channels becomes almost negligible on the global memory effect. Detecting the presence of classical correlation $\mu$ between multiple uses of quantum channels via the HSS tool becomes more challenging as the number of active qubits increases.

To further characterize the system behavior, we study the difference between the maximum and minimum values of HSS at a given dimensionless time $\tau$ and other fixed parameters. This difference $\delta$, characterizing the range of variation, is defined as 
\begin{equation}
    \delta=\max(\mathrm{HSS}(\mu=1,\tau,\cdots))-\min(\mathrm{HSS}(\mu=0,\tau,\cdots)),
\end{equation}
where "$\cdots$" denotes the other fixed parameters.
To do so, we demonstrate $\delta$ based on the number of qubits in Figs.~\ref{Multiqubit-Pauli-colored}{b}, \ref{Multiqubit-Pauli-Squeezed}{b}, and \ref{Multiqubit-depolarizing}{b}, respectively, for Pauli channels under colored pure dephasing, squeezed vacuum reservoir, and depolarizing under colored noise reservoir. In which $\delta$ decreases with an increasing number of qubits. It is anticipated that a large number of qubits, specifically beyond 10, will approach zero. This behavior indicates that the effects of the correlation strength between consecutive noise channels on the system are inhibited while increasing the number of qubits.

Further, for the multiqubit Pauli channel under a colored pure dephasing reservoir, we calculate the HSS for systems ranging from three to eight qubits. The dynamics of the HSS are depicted in Fig.~\ref{HSS_multiqubit_Pauli_pure_dephasing}. Increasing the number of active qubits, we find that the analysis given for the two-qubit model is also valid here. We also see that an increase in the number of qubits does not alter the time intervals in which non-Markovianity appears.

For the multi-qubit Pauli channel under a squeezed vacuum reservoir, the results align with those observed in the colored pure dephasing reservoir and the corresponding two-qubit model  (see Fig.~\ref{HSS_multiqubit_Pauli_squuezed}). Again, starting from a specific initial state, the time intervals at which non-Markovianity occurs remain unchanged despite an increase in the number of qubits. Similar observations can be made for multi-qubit depolarizing channels under a colored noise reservoir, as shown in Fig.~\ref{HSS_multiqubit_depolarizing}.

\textit{One of our most significant findings is that, assuming the qubits of a multipartite system are affected by independent or classically correlated local non-Markovian unital channels, the Hilbert-Schmidt speed (HSS) of the global system tends to zero as the number of qubits increases. This holds true regardless of the classical correlation strength between the noise channels. Our explicit results up to 8 qubits strongly suggest this general behavior for larger systems. Therefore, our findings indicate that the collective behavior of the multiqubit system inhibits the non-Markovian features of the overall system dynamics, as shown in Figs. \ref{Multiqubit-Pauli-colored}-\ref{HSS_multiqubit_depolarizing}.}

\section{Conclusion}\label{SecVII}
In correlated noisy channels, the global memory effect arises from both the temporal correlation created during evolution and the classical correlation between multiple uses of the quantum channel. Our study has shown that Hilbert Schmidt speed (HSS), a genuine witness, effectively captures the global memory effect in both unital and non-unital noisy channels. HSS is consistently sensitive to classical correlations, irrespective of the basis employed. While fluctuations in HSS dynamics can be attributed to the quantum non-Markovian evolution of the system, the amplification of these dynamics signals the presence of classical correlation between multiple uses of the quantum unital channel.

Furthermore, we investigated the influence of classical correlation on the non-Markovian memory effect of correlated noisy channels. For unital channels, classical correlation does not alter the duration of temporary revivals. However, in non-unital channels, increasing classical correlation varies the time intervals at which the non-Markovian effects appear. \textit{This makes Hilbert-Schmidt speed an effective tool to distinguish between unital and non-unital channels.
}
\par
Considering that the HSS is easily computable and does not require diagonalization, we extended our analysis to multiple qubits, extending up to eight qubits for unital noisy channels. The observed behavior in two-qubit noisy channels is replicated here: classical correlation amplifies the HSS dynamics, irrespective of the number of qubits. However, the impact of classical correlation between multiple applications of noisy channels on high-qubit systems is not dominant. As the number of qubits surpasses eight, HSS becomes less sensitive to classical correlations and primarily reflects the non-Markovian memory effect.

This study delves deep into the behavior of multiqubit systems under various noisy channels, particularly unital ones. \textit{As the number of qubits increases, the influence of classical correlations diminishes, suggesting a saturation point in the system capacity to encode additional classical information. }

A key result of our study is that when qubits in a multipartite system are influenced by either independent or classically correlated local non-Markovian unital channels, the Hilbert-Schmidt speed of the global system tends to zero as the number of qubits increases. This trend holds true irrespective of the classical correlation strength between the noise channels. Our comprehensive results, extending to systems with up to 8 qubits, strongly suggest that this behavior persists in larger systems. \textit{These findings underscore that the collective dynamics of multiqubit systems effectively suppress the non-Markovian features of the overall system evolution. } In our scenario where qubits are only entangled without any other direct interactions, the suppression of non-Markovianity can  be understood through the following points:

a)  The entanglement between qubits ensures that the system responds in a coordinated manner to local noises. Entanglement causes the qubits to respond collectively to local noises, resulting in a smoothing out of fluctuations through collective averaging and increased symmetry in the system dynamics. This coordinated response aligns more closely with Markovian behavior, and as the number of qubits increases, these uniform interactions become more pronounced, further diminishing the influence of memory effects.

b) Larger systems are more susceptible to pronounced decoherence, which accelerates the loss of coherence and dominates over non-Markovian characteristics. This strong decoherence obscures the signatures of non-Markovianity, such as information backflow, making them harder to detect and effectively suppressing their impact. Moreover, the scaling effects of decoherence, amplified by the interconnectedness of entangled systems, further diminish the observable non-Markovian behavior, ultimately leading to its suppression as the system size increases.
\par
Our results are crucial for advancing scalable quantum technologies, indicating that larger quantum systems may inherently resist classical noise correlations and mitigate collective non-Markovian effects. These findings pave the way for future research into the fundamental limits of quantum stability in complex systems. By leveraging the Hilbert-Schmidt speed as a diagnostic tool, researchers can further explore the resilience of quantum systems to environmental noise and develop strategies to enhance stability in practical quantum computing and communication applications. This study opens new avenues for understanding and optimizing the performance of large-scale quantum networks and devices.

\begin{acknowledgments}
	
	R.L.F. acknowledges support by MUR (Ministero dell’Università e della Ricerca) through the following projects: PNRR Project ICON-Q – Partenariato Esteso NQSTI – PE00000023 – Spoke 2 – CUP: J13C22000680006, PNRR Project QUANTIP – Partenariato Esteso NQSTI – PE00000023 – Spoke 9 – CUP: E63C22002180006. R.L.F. also thanks "Sistema di Incentivazione, Sostegno e Premialità della Ricerca Dipartimentale, Dipartimento di Ingegneria, Università di Palermo", project UNIPA D26-PREMIO-GRUPPI-RIC-2023.  H.R.J. wishes to acknowledge the financial support of the MSRT of Iran and Jahrom University.
\end{acknowledgments}

\section*{Author Contributions}
K.M. and S.N. equally contributed to this work. 
K.M.: Data curation (equal); Investigation (equal); Software (equal); Validation (equal); Visualization (equal); Writing – original draft (lead).
S.N.: Data curation (equal); Formal analysis (equal); Software (lead); Validation (lead); Visualization (equal); Writing – original draft (supporting).
H.R.J.: Conceptualization (equal); Formal analysis (equal); Investigation (supporting); Methodology (supporting); Project administration (lead); Supervision (supporting); Visualization (supporting); Writing – review and editing (equal).
R.L.F.: Conceptualization (equal); Formal analysis (equal); Project administration (supporting); Supervision (lead); Writing – review and editing (equal).

\section*{Data Availability Statement}
The data that support the findings of this study are available
from the corresponding author upon reasonable request.

\section*{Author Declarations }
The authors have no conflicts to disclose.

\appendix

\section{All Bases}\label{All Bases}
In this appendix, we provide an overview of various bases, including the standard, Bell, and Hadamard bases for two-qubit systems.

The standard computational basis is
\begin{equation*}
	\begin{split}
		\vert\psi_{00}\rangle =& \vert 0\rangle \otimes \vert 0\rangle, \\
		\vert\psi_{01}\rangle =& \vert 0\rangle \otimes \vert 1\rangle, \\
		\vert\psi_{10}\rangle =& \vert 1\rangle \otimes \vert 0\rangle, \\
		\vert\psi_{11}\rangle =& \vert 1\rangle \otimes \vert 1\rangle, \\
		\vert\psi_{S}\rangle =& \frac{1}{2} \left( e^{i \varphi} \vert \psi_{00}\rangle + \vert \psi_{01}\rangle + \vert \psi_{10}\rangle + \vert \psi_{11}\rangle \right).
	\end{split}
\end{equation*}\label{Standard Basis}

The Bell basis is
\begin{equation*}\label{Bell Basis}
	\begin{split}
		\vert\psi_+\rangle =& \frac{1}{\sqrt{2}} \left( \vert 0\rangle \otimes \vert 0\rangle + \vert 1\rangle \otimes \vert 1\rangle \right), \\
		\vert\psi_-\rangle =& \frac{1}{\sqrt{2}} \left( \vert 0\rangle \otimes \vert 0\rangle - \vert 1\rangle \otimes \vert 1\rangle \right), \\
		\vert\phi_+\rangle =& \frac{1}{\sqrt{2}} \left( \vert 0\rangle \otimes \vert 1\rangle + \vert 1\rangle \otimes \vert 0\rangle \right), \\
		\vert\phi_-\rangle =& \frac{1}{\sqrt{2}} \left( \vert 0\rangle \otimes \vert 1\rangle - \vert 1\rangle \otimes \vert 0\rangle \right), \\
		\vert\psi_B\rangle =& \frac{1}{2} \left( e^{i \varphi} \vert\psi_+\rangle + \vert\psi_-\rangle + \vert\phi_+\rangle + \vert\phi_-\rangle \right).
	\end{split}
\end{equation*}

The Hadamard basis is
\begin{equation*}\label{Hadamard Basis}
	\begin{split}
		\vert\psi_{+0}\rangle =& \frac{1}{\sqrt{2}} \left( \vert \psi_{00}\rangle + \vert \psi_{10}\rangle \right), \\
		\vert\psi_{+1}\rangle =& \frac{1}{\sqrt{2}} \left( \vert \psi_{01}\rangle + \vert \psi_{11}\rangle \right), \\
		\vert\psi_{-0}\rangle =& \frac{1}{\sqrt{2}} \left( \vert \psi_{00}\rangle - \vert \psi_{10}\rangle \right), \\
		\vert\psi_{-1}\rangle =& \frac{1}{\sqrt{2}} \left( \vert \psi_{01}\rangle - \vert \psi_{11}\rangle \right), \\
		\vert\psi_H\rangle =& \frac{1}{2} \left( e^{i \varphi} \vert\psi_{+0}\rangle + \vert\psi_{+1}\rangle + \vert\psi_{-0}\rangle + \vert\psi_{-1}\rangle \right).
	\end{split}
\end{equation*}

\section{Two-qubit correlated unital noisy quantum channels}\label{2Qubit_Pauli_all_basis}

In this appendix, we provide the explicit expressions of the HSS, in the Bell and Hadamard bases, for three different unital noisy channels considered in the manuscript. We denote the HSS in the Bell and Hadamard bases as $\mathrm{HSS}_{B}$ and $\mathrm{HSS}_{H}$, respectively.

\subsection{Two-qubit Pauli channel under a colored pure dephasing reservoir.} 

For a two-qubit Pauli channel under a colored pure dephasing reservoir, the two relevant HSS expressions in the Bell and Hadamard bases are given by, respectively,
   \begin{equation}
	\begin{split}
		\mathrm{HSS}_{B} =& \frac{1}{4} \sqrt{\left((1-\mu) \eta(\tau)^2 +\mu\right)^2 \cos^2(\phi) + \sin^2(\phi) + 2 \eta(\tau)^2}, \\
		\mathrm{HSS}_{H} =& \frac{1}{4} \left[((1-\mu) \eta(\tau)^2 + \mu)^2 + \eta(\tau)^2 (1 + \cos^2(\Phi))\right. \\ 
		&\left. + \sin^2(\phi)\right]^{1/2}.
	\end{split}
\end{equation} 
This indicates that the HSS is sensitive to $\mu$ regardless of the basis used.

\subsection{Two-qubit Pauli channel under a squeezed vacuum reservoir}

For a two-qubit Pauli channel under a squeezed vacuum reservoir, the HSS expressions in the Bell and Hadamard bases are obtained, respectively, as
\begin{widetext}
	\begin{equation}
		\begin{split}
			\mathrm{HSS}_{B} =& \frac{e^{-2 \gamma (\tau )}}{4} \left[ \cos^2(\Phi) \left( \mu \left( e^{2 \gamma (\tau )} - 1 \right) + 1 \right)^2  + e^{4 \gamma (\tau )} \sin^2(\Phi) + 2 e^{2 \gamma (\tau )}\right]^{1/2}, \\
				\mathrm{HSS}_{H} =& \frac{1}{4 \sqrt{2} (\cos (\Phi) + 2)^2} \bigg( 4 (\mu - 1)^2 e^{-4 \gamma (\tau )} (8 \cos (\Phi) + 3 \cos (2 \Phi) + 7) \\
			&+ 4 \left( 8 \mu^2 \cos (\Phi) + 3 \left( \mu^2 - 1 \right) \cos (2 \Phi) + 7 \mu^2 + 3 \right) \\
			&+ e^{-2 \gamma (\tau )} \left( (444 - 64 (\mu - 1) \mu ) \cos (\Phi) - 56 (\mu - 1) \mu + 327 \right) \\
			&+ e^{-2 \gamma (\tau )} \left( -24 (\mu - 3) (\mu + 2) \cos (2 \Phi) + 20 \cos (3 \Phi) + \cos (4 \Phi) \right) \bigg)^{1/2}.
		\end{split}
	\end{equation}
	\end{widetext}

\subsection{Two-qubit depolarizing channel under a colored noise reservoir}

The HSS functions for a two-qubit depolarizing channel under a colored noise reservoir in the Bell and Hadamard bases are given, respectively, by
\begin{widetext}
	\begin{equation}
		\begin{split}
			\mathrm{HSS}_{B} =& \frac{1}{4 \sqrt{2}} \bigg( (\Lambda - 1) \Lambda^2 (\mu - 1) (\Lambda (\mu - 1) - \mu - 1) \cos (2 \Phi) + 3 \Lambda^2 \left( (\Lambda (-\mu ) + \Lambda + \mu )^2 + 1 \right) \bigg)^{1/2}, \\
			\mathrm{HSS}_{H} =& \frac{1}{2 \sqrt{2} (\cos (\Phi) + 2)^2} \biggl( 8 \Lambda^2 \left( 7 (\Lambda (-\mu ) + \Lambda + \mu )^2 + 6 \right) \cos (\Phi) \\
			&+ 28 \Lambda^4 (\mu - 1)^2 - 50 \Lambda^3 (\mu - 1) \mu + \Lambda^2 (\mu (19 \mu + 6) + 28) + 3 \mu^2 \\
			&+ \Lambda^2 (\mu (49 \mu - 6) + 30) \cos^2(\Phi) \\
			&+ \Lambda^2 \left( (\Lambda (-\mu ) + \Lambda + \mu )^2 + 1 \right) \cos^3(\Phi) (\cos (\Phi) + 10) \\
			&+ \left( 40 \Lambda^4 (\mu - 1)^2 - 86 \Lambda^3 (\mu - 1) \mu - 3 \mu^2 \right) \cos^2(\Phi) \biggr)^{1/2}.
		\end{split}
	\end{equation}
\end{widetext}

\section{Two-qubit correlated non-unital noisy quantum channels}
For the amplitude damping channel, the HSS functions for the two Bell and Hadamard bases, respectively, are
\begin{widetext}
	\begin{equation}
		\begin{split}
			\mathrm{HSS}_{B} =& \frac{1}{4 \sqrt{2}} \bigg( (-2 G^3 - 4G) (\mu - 1) \mu + 2 \mu^2 + 2 G^8 (\mu - 1)^2 + G^4 (6 - 5 \mu) \mu + G^2 (\mu (7 \mu - 8) + 4) \\
			&+ G^2 \left(4 G^4 (\mu - 1)^2 - 2 G^6 (\mu - 1)^2\right) \cos (2 \Phi) + G^2 \left(G^2 (-( (\mu - 2) \mu + 2)) - 2 G (\mu - 1) \mu + \mu^2 \right) \cos (2 \Phi) \bigg)^{1/2}, \\
			\mathrm{HSS}_{H} =& \frac{1}{4 \sqrt{2} (\cos (\Phi) + 2)^2} \bigg( (40 G^4 - 6 G^4) (\mu - 1)^2 + G^2 (\mu - 1) (247 \mu - 327) + 367 \mu^2 + 4 \left(8 G^4 (\mu - 1)^2 - 222 G (\mu - 1) \mu \right) \cos (\Phi) \\
			&+ 4 \left(G^2 (\mu - 1) (95 \mu - 111) + 8119 \mu^2\right) \cos (\Phi) + (G (-\mu) + G + \mu)^2 (144 \cos (2 \Phi) + 20 \cos (3 \Phi) + \cos (4 \Phi)) \bigg)^{1/2}.
		\end{split}
	\end{equation}
\end{widetext}

\bibliography{ref}
\end{document}